\documentclass[prb,twocolumn,showpacs,preprintnumbers]{revtex4}
\usepackage{graphicx}
\usepackage{dcolumn}

\begin{document}

\title{Thermomagnetic Effects in Vortex Liquid:
\\ Transport Entropy Revisited}

\author{ A. Sergeev}
\email{asergeev@eng.buffalo.edu} \affiliation{ Research
Foundation, University at Buffalo, Buffalo, New York 14260}
\author{M. Yu. Reizer} \affiliation{5614 Naiche Rd.
Columbus, Ohio 43213}
\author{V. Mitin} \affiliation{Electrical Engineering
Department, University at Buffalo, Buffalo, New York 14260}

\begin{abstract}
Traditionally the Nernst and  Ettingshausen effects in the vortex
liquid are described in terms of the "transport entropy" of
vortices, $S_d$. According to current theories, the main
contribution to $S_d$ is originated from the electromagnetic free
energy, $F^{em}$, which includes kinetic and magnetic energy of
superconducting currents circulating around vortex cores. However,
this concept contradicts the London postulate, according to which
a supercurrent consists of macroscopic number of particles in a
single quantum state and does not transfer any entropy. Here we
resolve this contradiction and show that the transport entropy is
just ordinary thermodynamic entropy transferred by cores. Only in
this form the theory becomes simultaneously consistent with the
London postulate and the Onsager principle. The revised theory
explains measured temperature dependence $S_d$. The linear
increase of $S_d$ at low temperatures is determined by the entropy
of electrons in the core, then $S_d$ reaches a maximum at roughly
$T_c/2$ and then vanishes due to increase of the background
entropy.
\end{abstract}
\pacs{PACS   numbers: 71.10.-w} \maketitle

In recent years, extensive experimental studies of high-T$_c$
cuprates reveal a significant region of the phase diagram, in
which  a large Nernst effect and diamagnetism exist without the
long-range phase coherence.\cite{1,2,3} These important
observations are associated with the vortex liquid
formation.\cite{4,5} The central issue of this concept is the
Abrikosov's notion of the quantized flux line, which consists of a
normal core with the size of the coherent length, $\xi$, and
superconducting currents circulating around the core in the area
of the order of the magnetic penetration length, $\lambda$.
Counting the free energy of the vortex liquid from the level of
superconducting state, the free energy of cores $F^{core}$ may be
presented as a sum of the condensation energy and the energy
related to gradients of the order parameter.\cite{6} The free
energy of superconducting currents, defined as the electromagnetic
free energy $F^{em}$, includes the kinetic and magnetic energy of
the currents.\cite{6,7,8}

While it contradicts thermodynamics, current theories of
thermomagnetic vortex transport\cite{9,10,11,12,13} associate
$F^{em}$ with the thermal energy (for a review see Refs. 8 and
14). In other words, they attribute "transport entropy"
$S_{d}^{em}$ to the free energy of supercurrents. According to
this concept, the superconducting currents transfer the heat in
the Ettingshausen effect and create the net moving force
proportional to $-\nabla T$ in the Nernst effect. Moreover,
according to the current theories, the term $S_{d}^{em}$ 
significantly prevails over the core entropy. In limiting cases of
low and high magnetic field, $H-H_{c1}\ll H_{c1}$ and $H-H_{c2}\ll
H_{c2}$, the transport entropy $S_{d}^{em}$ was calculated using
the Ginzburg-Landau (GL) formalism.\cite{9,10,11} At the
intermediate fields, $S_{d}^{em}$ was obtained in the London
model,\cite{12} which was developed for extreme type-II
superconductors ($\xi \ll \lambda$), where cores are treated as
point singularities of the magnetic flux, i.e. $F^{core}=0$. The
London-type models are widely used for numerical studies of
thermomagnetic effects in high-T$_c$ cuprates.\cite{13}

Starting from famous works by Thomson (Lord Kelvin), the general
theory of thermoelectricity shows that the transport entropy of
thermal carriers should coincide with their thermodynamic entropy
counted from the background level.\cite{15} At the same time, the
supercurrents as any superfluid do not transfer thermodynamic
entropy.\cite{16,17} This is a direct consequence of the London
postulate, according to which the supercurrent is formed by
macroscopic number of particles moving coherently in a single
quantum state. Thus, the electromagnetic free energy $F^{em}$
related to macroscopic degrees of freedom does not consist the
entropy term.

In this work we revise the theory of thermomagnetic vortex
transport and resolve contradiction between the theory and the
London postulate. We show that the entropy $S_{d}$ is ordinary
thermodynamic entropy transferred solely by vortex cores. In
agreement with the London concept, the entropy of superconducting
currents is zero and they do not transfer the heat in the
Ettingshausen effect and do not produce the moving force in the
Nernst effect. In this way we reach an agreement with both the
Onsager principle and the London postulate.

In the Nernst effect the electric response is induced by a
transverse temperature gradient $\nabla T$. If the entropy $S_d$
moves from the area with the temperature $T$ to the area with the
temperature $T - \Delta T$, the ratio of the work produced by
thermal force, ${\bf f}_{th}\cdot \Delta {\bf r}$, to the thermal
energy, $T S_d$, is given by the Carnot efficiency $\Delta T/T$.
Therefore, the thermal force may be expressed as ${\bf f}_{th} =
-S_{d} \nabla T$.\cite{8,9,10,14} The thermal force ${\bf f}_{th}$
leads to the vortex motion with the velocity ${\bf v}_T = {\bf
f}_{th}/\eta $, where $\eta$ is the viscosity coefficient.
Magnetic flux of vortices $n \phi_0$ ($n$ is the vortex
concentration, $\phi_0$ is the flux quanta) generates the Nernst
{\it EMF}, which is $\tilde{\bf E}_N = n \overrightarrow{\phi}_0
\times {\bf v}_T/c$. Finally, the voltage signal in the open
circuit is given by
\begin{eqnarray}\label{N}
\tilde{\bf E}_N = {S_d \over c  \eta} \   \nabla T \times {\bf B},
\end{eqnarray}
where the magnetic field ${\bf B} = n \overrightarrow{\phi}_0$. In
the closed circuit the Nernst EMF generates the electric current
\begin{eqnarray}\label{Nj}
 {\bf j}^e  \equiv -\alpha \nabla T \times {\bf e}_B
 = \sigma_f \tilde{\bf E}_N = - {S_d \over c \phi_0}
 \ \nabla T \times {\bf e}_B,
\end{eqnarray}
where $\sigma_f=\eta / (\phi_0 B)$ is the flux-flow conductivity,
and ${\bf e}_B$ is the unit vector in the direction of ${\bf B}$.

In the Ettingshausen effect, the heat current is induced by the
transverse electric current ${\bf j}^e = \sigma_f {\bf E}$. The
current gives rise to the Lorentz force, ${\bf f}_L =({\bf j}^e
\times \overrightarrow{\phi}_0)/c $, which leads to the vortex
motion with the velocity ${\bf v}_L = {\bf f}_{L}/\eta $. The
thermal energy of a vortex is expressed in terms of the transport
entropy as $ \epsilon_{th} = T S_d $. Then, the heat current,
${\bf j}^h = n \epsilon_{th} {\bf v}_{L}$, may be presented as
\begin{eqnarray}\label{Ej}
{\bf j}^h \equiv \widetilde{\alpha}\ {\bf E} \times {\bf e}_B = {n
T S_d \over  c \eta} \ {\bf j}^e \times {\bf \phi}_0 = {T S_d
\over c \phi_0} \ {\bf E} \times {\bf e}_B.
\end{eqnarray}

The above description of thermomagnetic effects has been developed
by Stephen.\cite{9,10} Comparing Eqs. \ref{Nj} and \ref{Ej}, we
see that the Stephen formalism is in agreement with the Onsager
principle: $\widetilde{\alpha} = T \alpha$. This agreement is
reached by presenting both the thermal force $f_{th}$ and the
thermal energy $\epsilon_{th}$ via the transport entropy $S_d$.
However, after many years of extensive theoretical and
experimental research, the physical sense of $S_d$ and its
relation with ordinary entropy are still
unclear.\cite{8,9,10,11,12,13,14}

Previous theoretical works\cite{9,10,11,12,13} associate $S_d$
mainly with $F^{em}$. In his pioneering paper,\cite{9} Stephen
considered thermomagnetic vortex transport near $H_{c1}$, where
$B= n\phi_0 \approx 0$ and an interaction between vortices can be
neglected. In this case, $F^{em}$ per a vortex can be obtained in
the GL formalism,\cite{7,8,9,10}
\begin{eqnarray}\label{Uphi}
  F_{\phi}^{em} =  {\phi_0 \ H_{c1} \over 4\pi } =  \phi_0 | M(H_{c1})|=
  \Bigl({\phi_0 \over 4 \pi \lambda}\Bigr)^2 \ln {\lambda \over \xi},
\end{eqnarray}
where $ M = 4\pi H_{c1}$ is the magnetization. Stephen introduced
the transport entropy per vortex as $S_d^{St} = -\partial
F_{\phi}^{em}/
\partial T$ and from Eq. \ref{Uphi}
he obtained,\cite{9,10}
\begin{eqnarray}\label{St}
S_d^{St} =   -{\phi_0 \over 4\pi} {\partial H_{c1} \over
\partial T} =  - \phi_0 {\partial |M| \over
\partial T}= - {\partial \over \partial T} \Biggl({\phi_0^2 \over
16 \pi^2 \lambda^2} \ln {\lambda \over \xi}  \Biggr).
\end{eqnarray}
Note, the Stephen's approach could be easily generalized following
Ref. \cite{18} Within GL approach, Dorsey proved that the
electromagnetic free energy may be presented as $F^{em} = n
\phi_0|M|$. Thus, in the whole GL region
$F_{\phi}^{em}=\phi_0|M|$, so $S_d^{St}$ is given by $-
\phi_0\partial |M| / \partial T$.

Microscopic calculations of the Ettingshausen coefficient have
been done only near $H_{c2}$.\cite{11,19,20} Caroli and
Maki\cite{19} calculated $S_d$ using the TDGL formalism. However,
these calculations lead to the transport entropy, which diverges
at $T \rightarrow 0$. To get rid of this contradiction,
Maki\cite{20} suggested to complement the heat current with "the
thermodynamic thermal flux" due to magnetization: $j^h_{mag} =
-{\bf E} \times {\bf M}$. Near $T_c$, in the impure limit, $\ell
\ll \xi$ ($\ell$ is the electron mean free path), the corrected
result is given by
\begin{eqnarray}\label{Maki}
S_d^{Mk} = {\phi_0 |M| \over T}= {\phi_0\over 4\pi T}
{H_{c2}-H\over \beta_A(2\kappa^2-1)},
\end{eqnarray}
where $\beta_A$ is the geometrical factor. While Eq. \ref{Maki} is
widely used to fit experimental data, it is well-known\cite{8}
that, the magnetization correction $j^h_{mag}$ leads to violation
of the Onsager principle. Finally, Troy and Dorsey\cite{11}
reproduced Eq. \ref{Maki} by associating the thermal energy of a
vortex $\epsilon_{th}=T S_d$ with the electromagnetic free energy
$F_{\phi}^{em} = \phi_0|M|$.

As we can see, the results of the phenomenological and microscopic
theories are not in agreement: the Stephen's entropy (Eq.
\ref{St}) is proportional to the temperature derivative of $M$,
while the Maki entropy (Eq. \ref{Maki}) is proportional to $M/T$.
It is even more important that the conclusions of the
phenomenological theory and the interpretation of the microscopic
theory in terms of $F^{em}$ contradicts the London postulate.

Now we will show that the entropy of superconducting currents in
fact is zero and, in agreement with the Onsager principle, the
superconducting currents do not contribute to the Nernst effect as
well. First, we would like to note that in the Stephen theory
\cite{9} the nonzero entropy of supercurrents has been obtained
due to misinterpretation of thermodynamic relations. If $S=0$, the
free energy, $F^{em}=U^{em}-TS$, is equal to the internal energy,
i.e. $U^{em} = F^{em}$ and $U_{\phi}^{em}=F_{\phi}^{em} = \phi_0
|M|$  Then, the entropy can be again expressed via thermodynamic
relation as the temperature derivative of  $F^{em}$ (without the
energy of magnetic field $H^2/8\pi$) {\it at the constant
magnetization,}\cite{21} i.e.
\begin{eqnarray}\label{Sc}
S^{em} = -\Bigl({\partial F_{\phi}^{em} \over \partial T}\Bigr)_M
= - n \phi_0 \Bigl({\partial M \over \partial T}\Bigr)_M =0.
\end{eqnarray}
Of course, this consideration does not add anything new beyond the
London's postulate. It just shows that, in fact, the temperature
derivative in Eq. \ref{St} should be calculated at constant M,
which results in zero entropy. We also see that, contrary to Ref.
\cite{20}, the superconducting magnetization currents do not
participate in the heat transfer and, therefore, consistent
microscopic calculations of the heat current do not require any
artificial thermodynamic corrections due to
magnetization.\cite{22,23}

Now we consider the Nernst effect. Let us start with
noninteracting vortices near $H_{c1}$. Taking into account the
temperature dependence of the vortex energy
$U_{\phi}^{em}=F_{\phi}^{em}$ (Eq. \ref{Uphi}), the thermal force
in the Nernst setup can be calculated as
\begin{eqnarray}\label{fth}
{\bf f}_{th} &=& -{\partial U_{\phi}^{em} \over \partial {\bf r} }
= - {\partial \over \partial T} \Biggl({\phi_0^2 \over 16 \pi^2
\lambda^2} \ln {\lambda \over \xi}  \Biggr) \nabla T \nonumber \\
&=& - \phi_0 {\partial |M| \over \partial T} \ \nabla T.
\end{eqnarray}
As it is shown in Fig. 1 (a), the thermal force is directed from
cold to hot area, because the vortex energy $U_{\phi}(T)$
decreases when $T$ increases. Note, that our thermal force
obtained under the assumption of $S^{em} =0$ (Eq. \ref{fth}) and
the thermal force introduced by Stephen\cite{8,9,10} have the same
value, but opposite directions.

To satisfy the Onsager principle, the thermal force should be
balanced by another force. The additional force overlooked in all
previous works originates from the magnetization currents  in the
presence of $\nabla T$,\cite{23,24}
\begin{eqnarray}\label{jT}
{\bf j}^e_{\nabla T} =  c \nabla \times {\bf M} (T)= c \nabla T
\times {\partial {\bf M} \over \partial T}.
\end{eqnarray}
As it is shown in Fig. 1 (b), the current ${\bf j}^e_{\nabla T}$
leads to the Lorentz force, which acts on a single vortex in the
direction perpendicular to $\nabla T$,
\begin{eqnarray}\label{flor}
{\bf f}_L &=& {1\over c} \ {\bf j}^e_{\nabla T} \times
\overrightarrow{\phi}_0
 = - \Biggl(
\overrightarrow{\phi}_0 \cdot {\partial {\bf M} \over
\partial T}\Biggr) \ \nabla T \nonumber \\   &=&  {\phi}_0
{\partial |M| \over
\partial T} \ \nabla T.
\end{eqnarray}
The Lorentz force ${\bf f}_L$ is directed from hot to cold area
(Fig. 1 (b)). Thus, Eqs. \ref{fth} and \ref{flor} show that the
total moving force acting on vortex supercurrents is zero.

Employing the Dorsey result,\cite{18} the above conclusion can be
generalized for interacting vortices ($1/\lambda^2 < n < 1/\xi^2$)
and even for overlapping cores ($n \sim 1/\xi^2$). As we discussed
above, the TDGL formalism leads to the general expression for the
electromagnetic free energy $F^{em}=U^{em}= nU_{\phi}= n \phi_0
|M|$, i.e. vortices can considered as independent elementary
excitations with energy $U_{\phi}^{em}=\phi_0 |M|$. Then, Eqs.
\ref{fth} and \ref{flor} expressing the thermal and Lorentz forces
through the magnetization are also valid for the entire mixed
state and, therefore, the balance of forces is universal.
Moreover, the Dorsey result and the above proof are applicable to
any pairing.

We have shown that thermomagnetic effects are absent as long as we
limited our consideration by $F^{em}$. To get nonzero effects, we
should take into account contributions of normal electrons, i.e.
$F^{core}$. The transport entropy is an ordinary thermodynamic
entropy counted from a background. If  vortex cores do not overlap
each other, i.e. $n \xi^2 \ll 1 $, the background is "pure"
superconducting and, therefore, the transport entropy is
determined by the condensation energy, $H_c^2/8\pi$, in the core
area, which is $\sim \pi \xi^2$. Thus, the transport entropy per a
vortex is
\begin{eqnarray}\label{entr}
S_d^{core} (T) \simeq - \pi \xi^2 \ {\partial  \over \partial T}
{H_c^2(T) \over 8 \pi}.
\end{eqnarray}
Note that close to $H_{c2}$, i.e. in the magnetic field $H \simeq
B \simeq \phi_0/\xi^2$, the background is formed by cores of other
vortices and Eq. \ref{entr} is inapplicable. Here, the transport
entropy $S_d$ decreases due to overlapping of vortex cores and
goes to zero at $H_{c2}$. Self-consistent description of the
narrow region near $H_{c2}$ requires microscopic consideration.

The exact results for $S_d$ can be found from the GL formalism in
the limit of large $\kappa$. In this case the condensation energy
and the transport entropy are\cite{7,8}
\begin{eqnarray}\label{}
F_{\phi}^{core} = a \Bigr({ \phi_0 \over 4 \pi \lambda}\Bigl)^2, \
\ \ S_d^{core} (T) = {\partial F_{core} \over \partial T}.
\end{eqnarray}
In the original paper by Abrikosov, the constant $a$ was found to
be $\sim 0.08$,\cite{7} then Hu\cite{25} corrected its value to
0.497. Comparing with Eq. \ref{St}, we see that in this limiting
case the correct value of $S_d$ is approximately $2
\ln(\lambda/\xi)$ times smaller than that predicted by
Stephen.\cite{9}

Now let us analyze the measured temperature dependence of $S_d$.
Detailed numerical analysis\cite{26} shows that at moderate
temperatures $ 0.2 \leq T/T_c \leq 0.9 $ the radius of the vortex
core $\xi_1$, defined by fitting the pair potential $\Delta(r)$ by
an expression $\Delta(r)=\Delta_0\tanh(r/\xi_1)$, just weakly
depends on temperature. Therefore, according to Eq. \ref{entr} the
temperature dependence of $S_d$ is mainly determined by the
dependence $H_{c}(T) \propto 1-(T/T_c)^2$, so $S_d$ is
proportional to $(T/T_c)[1-(T/T_c)^2]$ and has a smooth maximum at
$T \simeq 0.6 T_c$. In Fig. 2 we compare the above conclusions
with the temperature dependence of the transport entropy
determined by Solomon and Otter\cite{27} from the Ettingshausen
effect in InPb films. As seen, we get a good agreement with the
data. The linear increase of $S_d$ at low temperatures is
determined by the entropy of electrons in the core, then $S_d$
reaches a maximum and vanishes due to increase of the background
entropy. Using parameters InPb alloy\cite{27}, we evaluate the
maximum of $S_d(T)$ is $1.2 \cdot 10^{-7}$ erg/cm K, while the
experiment gives $2 \cdot 10^{-7}$ erg/cm K. Thus, the proposed
model provides a simple explanation of the nonmonotonic
temperature dependence of $S_d$ in ordinary superconductors.

Origin of giant thermomagnetic effects in high-$T_c$ cuprates is a
key point for understanding of the nature of the ground state in
these materials.\cite{1,2,3,4,5} We have shown\cite{23} that
thermomagnetic coefficients in the Fermi liquid are always
proportional to the square of the particle-hole asymmetry (PHA).
It means that the giant effects cannot be explained by the
interaction effects in the Fermi liquid, e.g. by superconducting
fluctuations\cite{28,29}. The explanation requires strong PHA,
e.g. the Fermi-surface reconstruction due to the spin density wave
gap,\cite{30} or a non-Fermi liquid state such as the vortex
liquid\cite{2,3}. The formation of the vortex liquid is associated
with the 3D analog of the Kosterlitz - Thouless
transition.\cite{4,5} While the thermodynamics of this unusual
phase is still under debates and various models are proposed, our
conclusion that the transport entropy is the thermodynamic entropy
transferred by cores is fully applicable to any vortex model.

In summary, we have shown that the superconducting currents
circulating around cores do not contribute to $S_d$, i.e. the
supercurrents do not transfer the thermal energy in the
Ettingshausen effect and do not produce the moving force
proportional to $-\nabla T$ in the Nernst effect. Only this
approach is consistent with thermodynamics of irreversible
processes (i.e. the Onsager relation and the third law of
thermodynamics) and the London postulate. According to the London
postulate, any superconducting currents, including superconducting
fluctuation currents, do not transfer entropy and thermal energy.
It is surprising that all recent papers related to the fluctuation
region above the mean-field transition temperature state opposite
and insist on the heat transfer by fluctuation magnetization
currents (i.e. see Refs. \cite{28,29} and our comment\cite{31}).
For the vortex liquid, we have shown that the transport entropy of
vortices is just ordinary thermodynamic entropy of cores counted
from the background entropy. In this way we have connected
thermomagnetic transport with thermodynamics. Our theory provides
natural explanation of nonmonotonic dependence $S_d(T)$ in
ordinary superconductors. It can be easy generalized for various
models, which were recently suggested for the vortex liquid in
cuprates.

We are grateful to I. Aleiner, A. Gurevich, and N. Kopnin for
valuable discussions.

\begin{figure}

\caption{Balance of two forces acting on the superconducting
currents: (a) ${\bf f}_{th}$ is the thermal force (Eq. 8), (b)
${\bf f}_L$ is the Lorentz force due to the magnetization currents
(Eq. 10).}

\caption{The temperature dependence of the transport entropy:
theory (solid line) and data from Ref. 27.}
\end{figure}

\end{document}